\documentstyle[floatflt,aps,epsf]{revtex}  

\newcommand{\ff}{fragmentation function}
\newcommand{\sv}{scaling violation}
\newcommand{\mul}{multiplicit}
\newcommand{\cacf} {\ifmmode{C_A/C_F}       \else{$C_A/C_F$}       \fi}
\newcommand{\epem} {\ifmmode{e^+e^-}        \else{$e^+e^-$}       \fi}
\newcommand{\qqbar} {\ifmmode{q\bar{q}}     \else{$q\bar{q}$}       \fi}
\newcommand{\tje} {three jet event}
\newcommand{\mean}[1]{\ifmmode{{\langle{#1}\rangle}}
                      \else{$\langle{#1}\rangle$} \fi}
\newcommand{\eref}[1]{\mbox{{\sc Eq.~}\ref{#1}}}

\newcommand{\fref}[1]{\mbox{{\sc Fig.~}\ref{#1}}}
\newcommand{\lref}[1]{\cite{#1}}
\newcommand{\chref}[1]{\mbox{{\sc Sec.~}\ref{#1}}}
\newcommand{\eps}{{\ifmmode \varepsilon        \else $\varepsilon$\fi}}
\renewcommand{\th} {{\ifmmode \vartheta          \else $\vartheta$\fi}}
\newcommand{\into} {{\ifmmode \rightarrow       \else $\rightarrow$\fi}}
\newcommand{\BC}{\begin{center}}
\newcommand{\EC}{\end{center}}
\newcommand{\BE}{\begin{equation}}
\newcommand{\EE}{\end{equation}}
\newcommand{\BA}{\begin{array}}
\newcommand{\EA}{\end{array}}
\newcommand{\BI}{\begin{itemize}}
\newcommand{\EI}{\end{itemize}}
\newcommand{\BF}{\begin{figure}}
\newcommand{\EF}{\end{figure}}
\newcommand{\BT}{\begin{table}}
\newcommand{\ET}{\end{table}}
\newcommand{\BTB}{\begin{tabular}}
\newcommand{\ETB}{\end{tabular}}
\newcommand\BM{\begin{minipage}}
\newcommand\EM{\end{minipage}}

\newcommand\delphi{{\sc Delphi}}
\newlength{\wi}   \wi 8cm
\newlength{\fwi} \fwi 0.95\wi
\begin{document}        

\baselineskip 14pt
\title{Comparing Gluon to Quark Jets with {\sc Delphi}}
\author{Oliver Klapp, Patrick Langefeld}
\address{Univ. Wuppertal, FB 8, Gau{\ss}str. 20, 42097 Wuppertal, Germany\\
{\sc Delphi} Collaboration\\
E-mail: Klapp@WHEP.uni-wuppertal.de}
%
\maketitle              

\begin{abstract}        
This is a summary of the latest results of the {\sc Delphi}
Collaboration on the properties of identified quark and gluon jets.
It covers the measurement of 
the \ff s of gluons and quarks and their \sv\
behaviour as well as an analysis of the scale dependence of the multiplicities
in
gluon and quark jets. Further, a precision measurement of \cacf\ from the
multiplicities in symmetric \tje s is discussed.
\
\end{abstract}   	

\section{Introduction}\label{sec:Introduction}

In QCD, the three fundamental splittings of the
two types of colour charged fields participating in
the strong interaction (quarks~($q$) and gluons~($g$))
are
$q\to qg$,
$g\to gg$, and
$g\to \qqbar$.
The corresponding {\em splitting kernels}, which describe both the kinematics
and the relative strengths of these splittings, are proportional to
the {\em colour factors}
$C_F=\frac{4}{3}, C_A=3$, and $T_R=\frac{1}{2}n_F$, respectively,
where $n_F$ is the number
of active quark flavours in $g\to \qqbar$ decays.
The values for these colour factors originate directly from
the group $\cal{SU}$(3) underlying QCD so that precise measurements of these
couplings validate QCD as the fundamental theory for strong interactions.
In the sensible range of $n_F=3\dots 5$, the
ratio of the splitting kernels
$S_{(g\to gg+g\to\qqbar)} / S_{(q\to qg)}$
is nearly constant (2.0 - 2.3) and exactly
\cacf=2.25 in limit of large energy transfers. This leads to the expectation of
a \cacf\ times higher probability of gluon bremsstrahlung in gluon jets with
respect to quark jets. This ratio should be visible for observables
which are proportional to the splitting probabilities of gluons and
quarks\lref{l:brodsky_gunion}.

Unfortunately, quarks and
gluons are no free particles.
Therefore one has to use jets of gluons and quarks in three jet events as the
best approximation of the tree level graphs.
Beyond the necessity to rely on the assumption of LPHD 
({\bf L}ocal {\bf H}adron {\bf P}arton {\bf D}uality) which states that the
parton properties are preserved to the hadron level which is then
clustered to jets,
this approach is further limited by
the influence of interference effects in the event.
Moreover the assignment of particles to jets is somewhat arbitrarily.

Further, the jet finding algorithms introduce
ambiguities when assigning particles to jets.

\section{Experimental access to gluon and quark jets}\label{sec:exp}

\setcounter{figure}{1}
\begin{floatingfigure}[l]{6cm}
\centerline{\epsfxsize 6cm \epsfbox{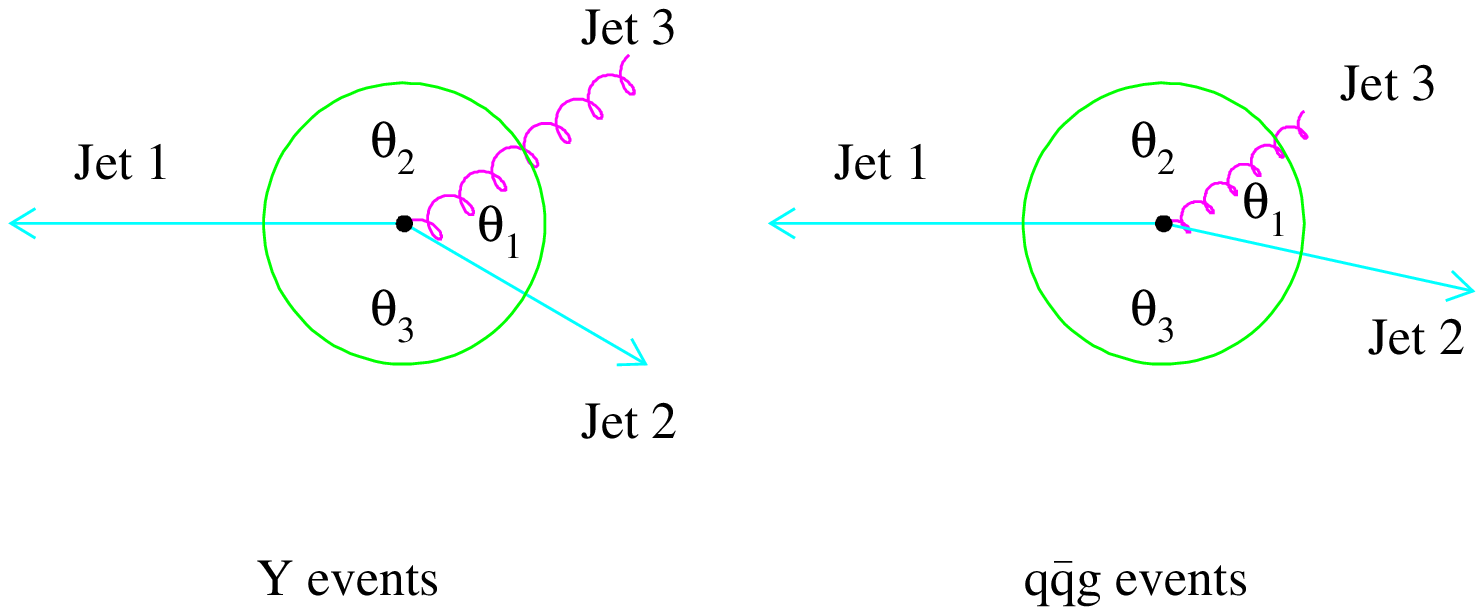}}
\centerline{
\parbox[b]{6cm}{
 FIG. 1. Symmetric and asymmetric event topologies
 } 
}
\end{floatingfigure}

Gluon jets were originally identified in symmetric
\tje s ({\em Y events})\lref{l:p:spain,l:dip:oli,l:dip:martin}.
In these events the most energetic jet is
excluded from the analysis, as the rate of gluon induced leading jets
is rather low.
Due to the symmetric event topology ($\theta_2\sim\theta_3$,
see~{\sc Fig.~}1
), the low energy jets are expected to be directly
comparable.
By identifying one of these jets as a $b$-quark
jet using impact parameter techniques,
the remaining jet is identified indirectly as a gluon jet.
The properties of a
comparable $udsc$-quark jet sample can then be obtained from non-$b$
events, where the gluon properties are eliminated from 
by subtraction techniques~\cite{l:dip:oli,l:dip:martin}.

Recently this technique has also been extended to non-symmetric
events\lref{l:dip:martin} (see~{\sc Fig.~}1
). In this case one relies on the
quark/gluon composition of jets as predicted by the three jet matrix
elements taken from Monte Carlo simulations.
This technique improves the available statistic
and gives access to a wider range of energy scales,
but requires a criterion for the selection of comparable
gluon and quark jets. These jets are 
obtained from events with different topologies.
The comparison of jet properties obtained for this event sample\footnote{
In the following called {\em asymmetric} event sample though containing
all Y events as well.}
and for Y events helps in finding a suitable scale to classify the jets with.
One yields about 20,000 identified gluons in Y events
and about 100,000 in the asymmetric events.

\section{Jet scales}
\label{sec:energy}

\begin{figure}[t]
\wi 0.48\textwidth
\fwi 0.46\textwidth
\begin{minipage}[t]{\wi}
\centerline{\epsfxsize \fwi \epsfbox{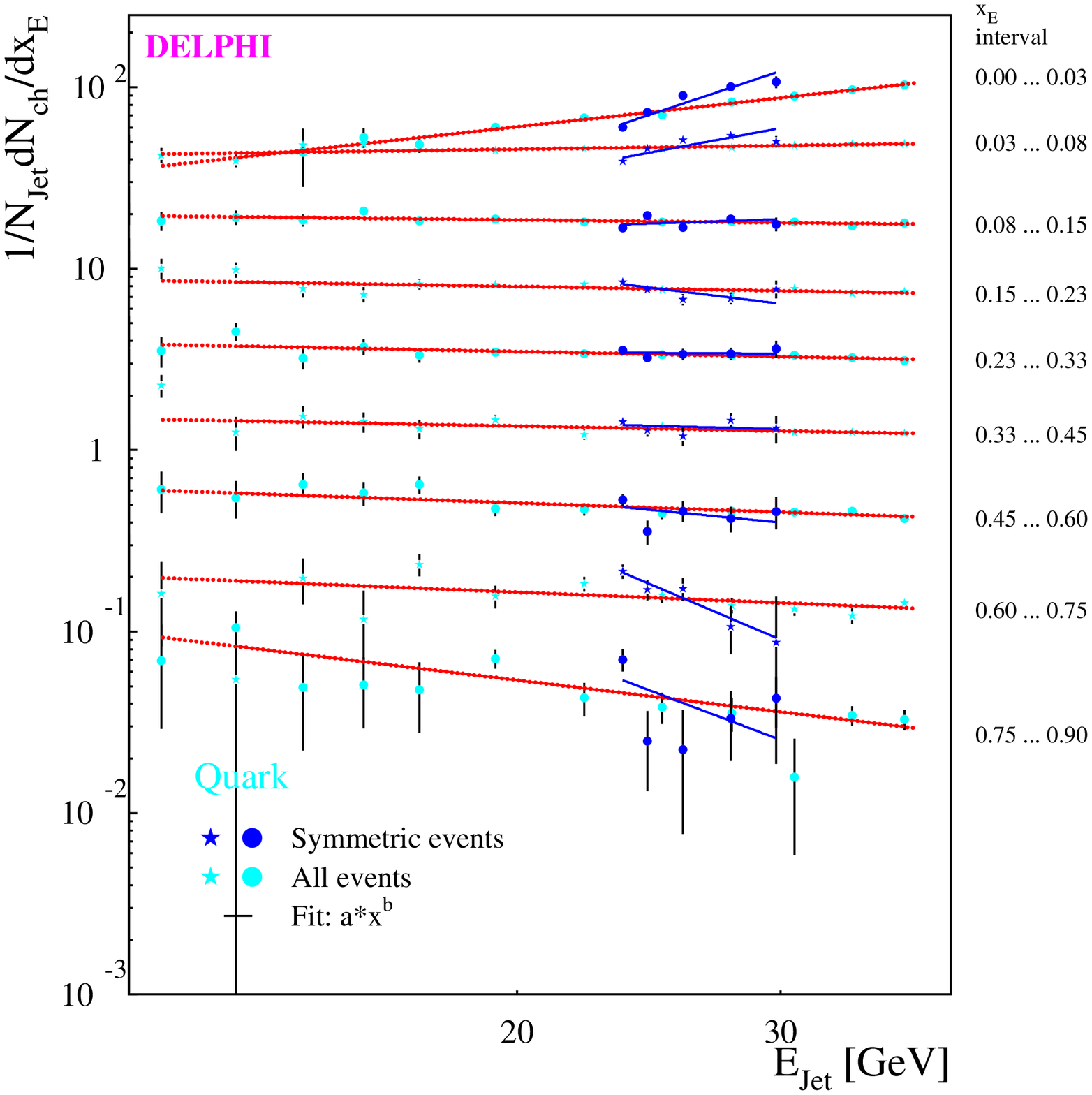}}
\caption{Quark frag. func. vs. $E_{Jet}$. }
\label{f:fan-q-1}
\end{minipage}
\hfill
\begin{minipage}[t]{\wi}
\centerline{\epsfxsize \fwi \epsfbox{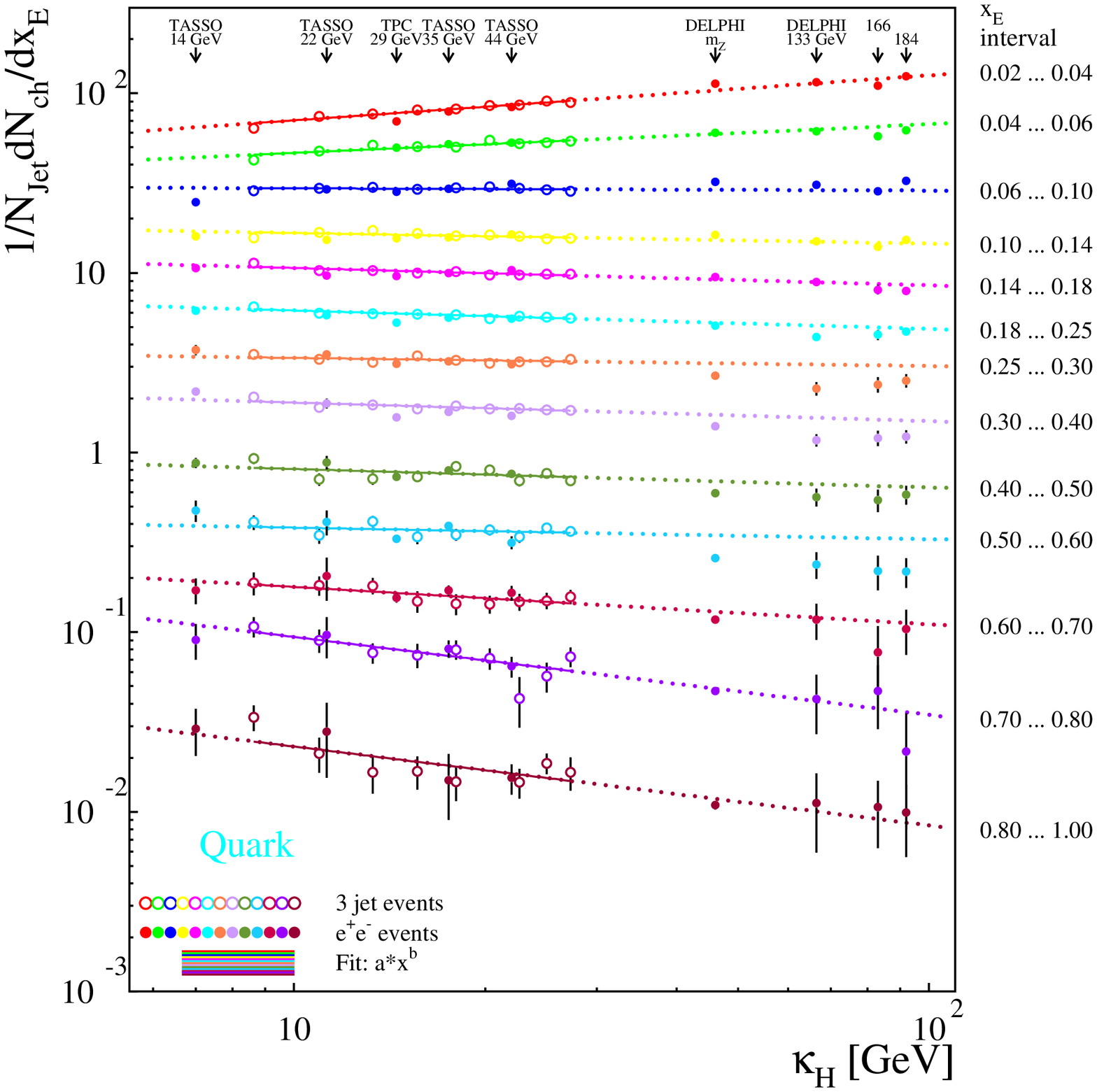}}
\caption{Quark frag. func. vs. $\kappa_H$}
\label{f:fan-low}
\end{minipage}
\end{figure}

The relevant scale for the jet evolution is not just the jet
energy. In~\fref{f:fan-q-1}, the \ff s of quark jets as a
function of their energy are shown. The {\em rows} in
this figure correspond to data taken in the same $x_E$ intervals.
The much more pronounced scaling violation pattern of jets
obtained from symmetric events ($E_{Jet}\sim22 GeV\dots 29 GeV$)
compared to jets from asymmetric events
clearly disfavours the jet energy as the relevant scale.
Soft radiation is limited to cones given by the opening angles between the
jets. This motivates transverse momentum scales. 
The {\em hardness}
$\kappa_H=E\sin{\theta_1/2}$
is a better choice\lref{l:khoze_ochs}, as it
accounts for the limited phase space available for
gluon radiation due to the interference of radiated gluons in 
the event\footnote{This definition corresponds to 
the {\bf beam} energy of an $\epem\to\qqbar$ event. Often also the definition
$\kappa = 2E\sin{\theta/2} \sim E\theta$ is used.}.
\fref{f:fan-low} shows the good agreement of
light quark jets in \tje s with jets in
\epem
annihilations from {\sc Petra} energies\lref{l:frag_tasso} to the highest
LEP energies\lref{l:passon} in normalization and slopes using $\kappa_H$ for
the first and $E_{beam}$ for the latter. This agreement confirms
the interpretation of $\kappa_H$ as a valid scale for jet evolution in
\tje s.
In multi-jet events several scales may be relevant; in so far
the usage of $\kappa_H$ as a (single) scale is an approximation.
Another possible choice is the scale $p_1^T$, introduced in~\chref{sec:khoze}.

\section{Results}\label{sec:results}

Recently, several properties of quark and gluon jets 
which can be expected to be sensitive to 
\cacf , such as the scaling
violation of quark and gluon jets\lref{l:vanc:d_scaling}
and the scale dependence of the
quark and gluon jet multiplicities\lref{l:vanc:d_multi} have been analysed 
using data collected with the \delphi\ detector at {\sc LEP}.

\subsection{Gluon and Quark Fragmentation Functions}\label{sec:glufrag}

The gluon and quark fragmentation functions $D_{g,q}(x_E,\kappa_H)$ 
are measured in
dependence of different values of $\kappa_H$ at a fixed CMS energy
(see~\chref{sec:energy} for a short discussion of jet scales).

\wi 0.48\textwidth
\fwi 0.46\textwidth
\begin{figure}[t]
\begin{minipage}[t]{\wi}
\centerline{\epsfxsize \fwi \epsfbox{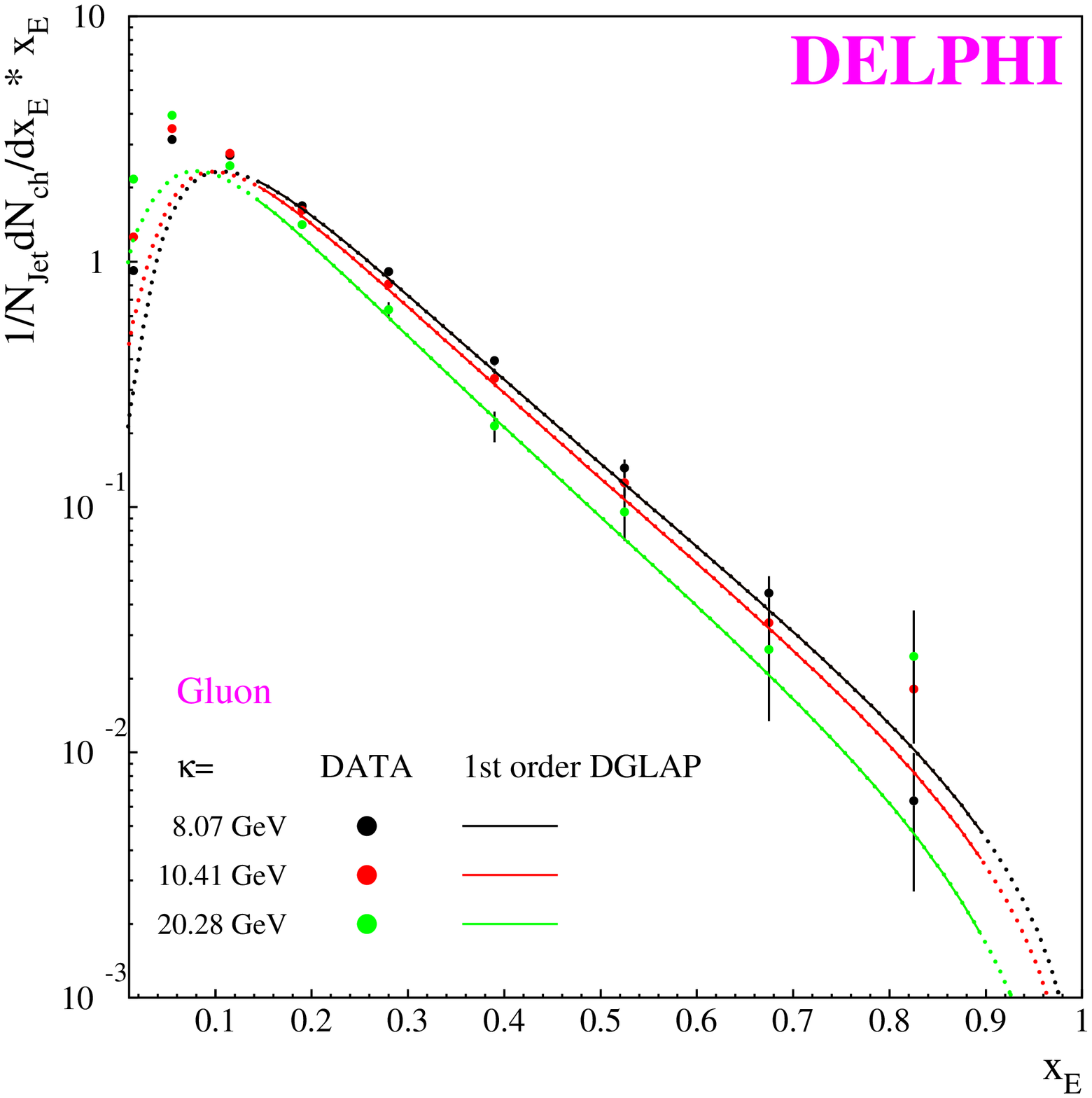}}
\caption{Gluon fragmentation functions for different values of $\kappa_H$}
\label{f:sc-fanx-g}
\end{minipage}
\hfill
\begin{minipage}[t]{\wi}
\centerline{\epsfxsize 8cm \epsfbox{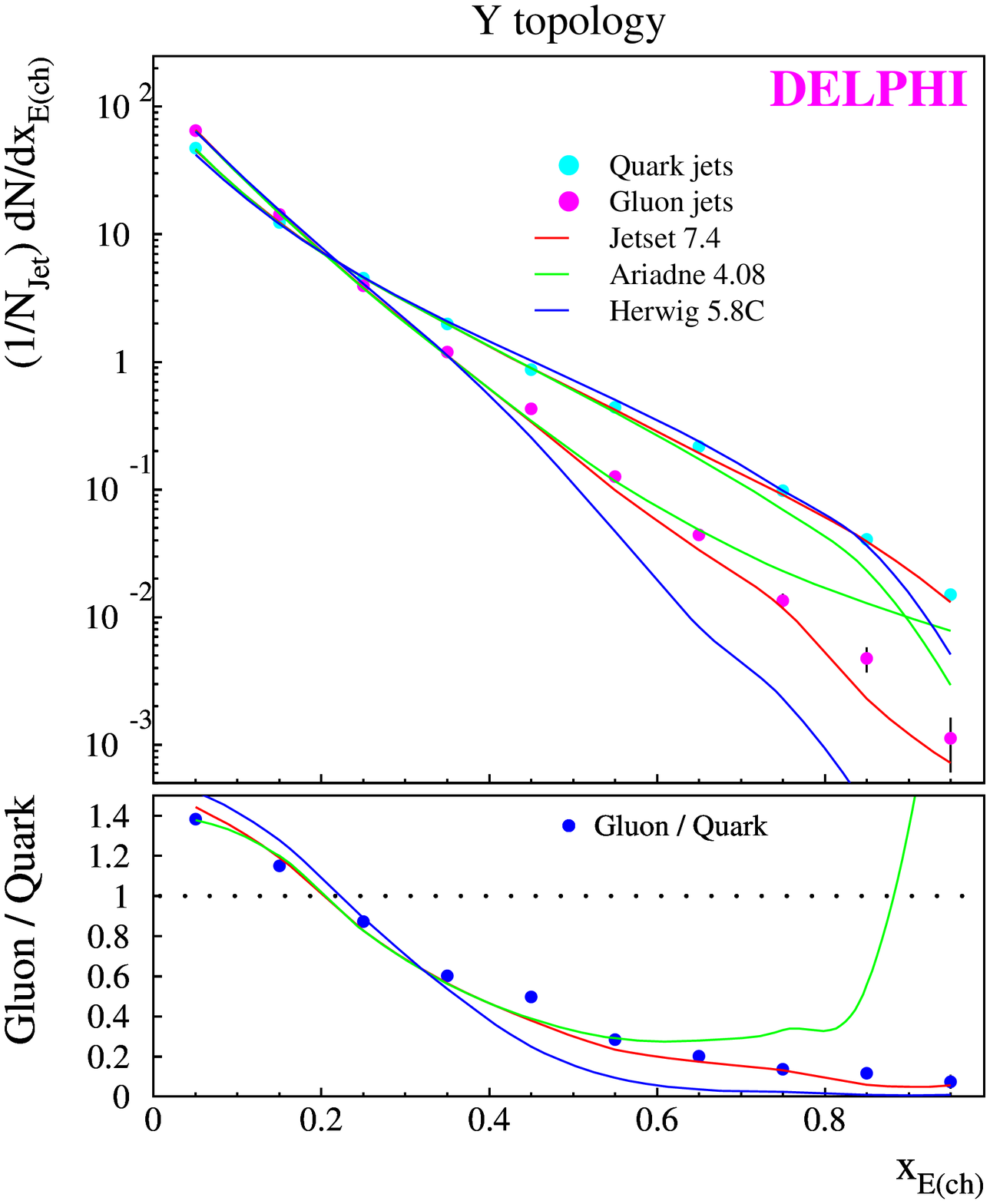}}
\caption{Gluon and quark fragmentation function at
a fixed topology, $\theta_{2}, \theta_{3} \in [150^\circ \pm 15^\circ]$ ),
compared to different Monte CArlo generators}
\label{f:xech}
\end{minipage}
\end{figure}

The predictions are derived from the numerical solution of the DGLAP evolution
equation in first order~\lref{l:dglap}. The following ansatz has been used to parameterize
the fragmentation
function at a fixed scale $\kappa_0$ to start the evolution:
\begin{eqnarray}
  D_p^{g,q}(x_E) = p^{g,q}_3 \cdot x_E^{p^{g,q}_1} \cdot 
         (1-x_E)^{p^{g,q}_2}
      \cdot \exp{(-p^{g,q}_4 \cdot \ln^2{x_E})} \, .
\label{gl_dfrac}
\end{eqnarray}
The parameters $p_i^{q,g}$, $\Lambda_{QCD}$ and the colour factor $C_A$ are
fitted simultaneously.

The softening of the fragmentation functions with increasing
$\kappa$ is observed. This effect is more
pronounced for gluon jets than for quark jets.
Fig.~\ref{f:xech} compares the gluon and the quark fragmentation function at
a fixed topology.

\subsection{Scaling violation of the gluon and quark \ff s}
\label{sec:scaling}

\begin{figure}[t]
\begin{minipage}[t]{\wi}
\centerline{\epsfxsize \fwi \epsfbox{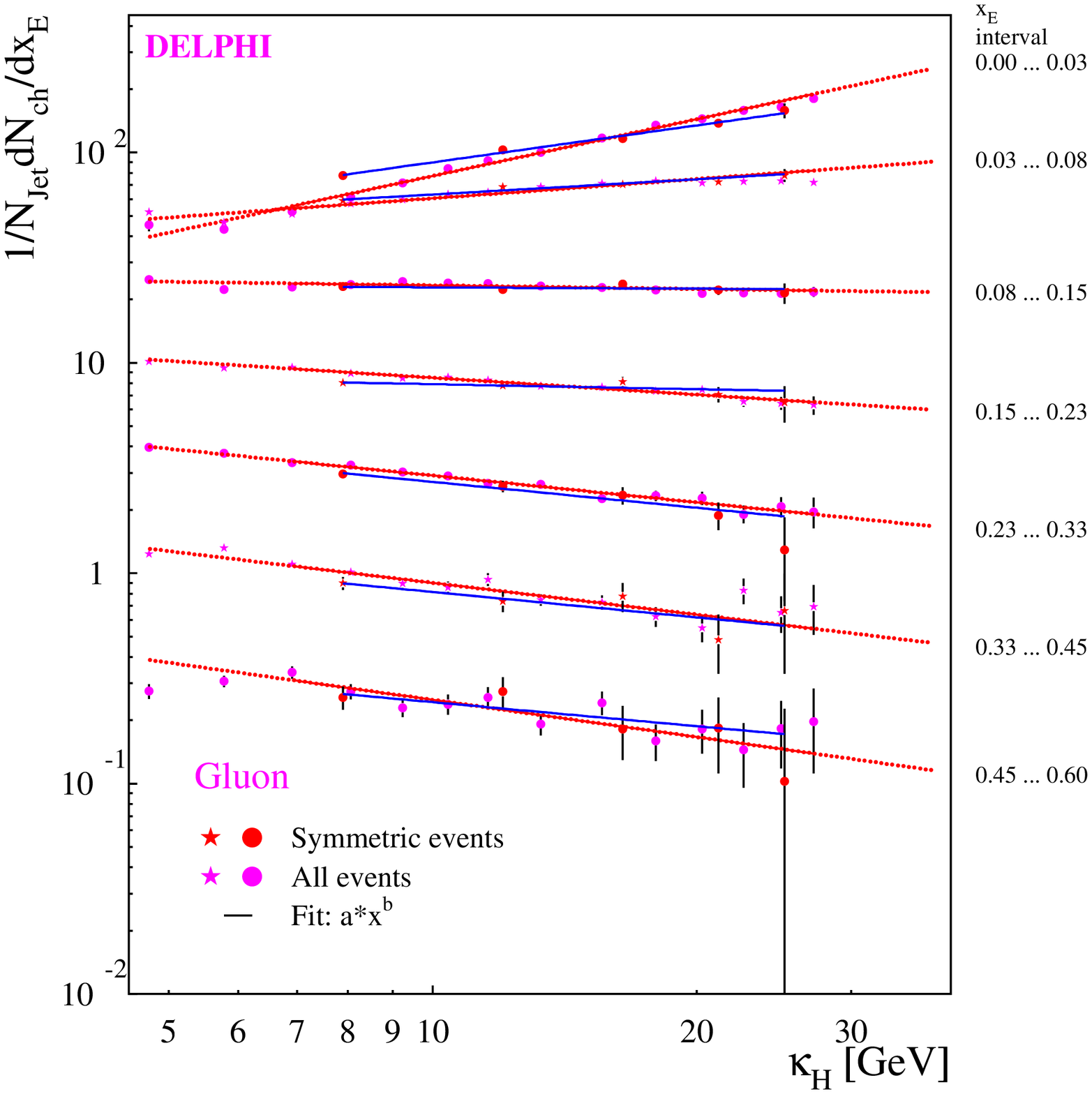}}
\caption{Gluon \ff s as a function of jet scale $\kappa_H$. Superimposed are
jets from both symmetric and asymmetric events}
\label{f:fan-g-2}
\end{minipage}
\hfill
\begin{minipage}[t]{\wi}
\centerline{\epsfxsize \fwi \epsfbox{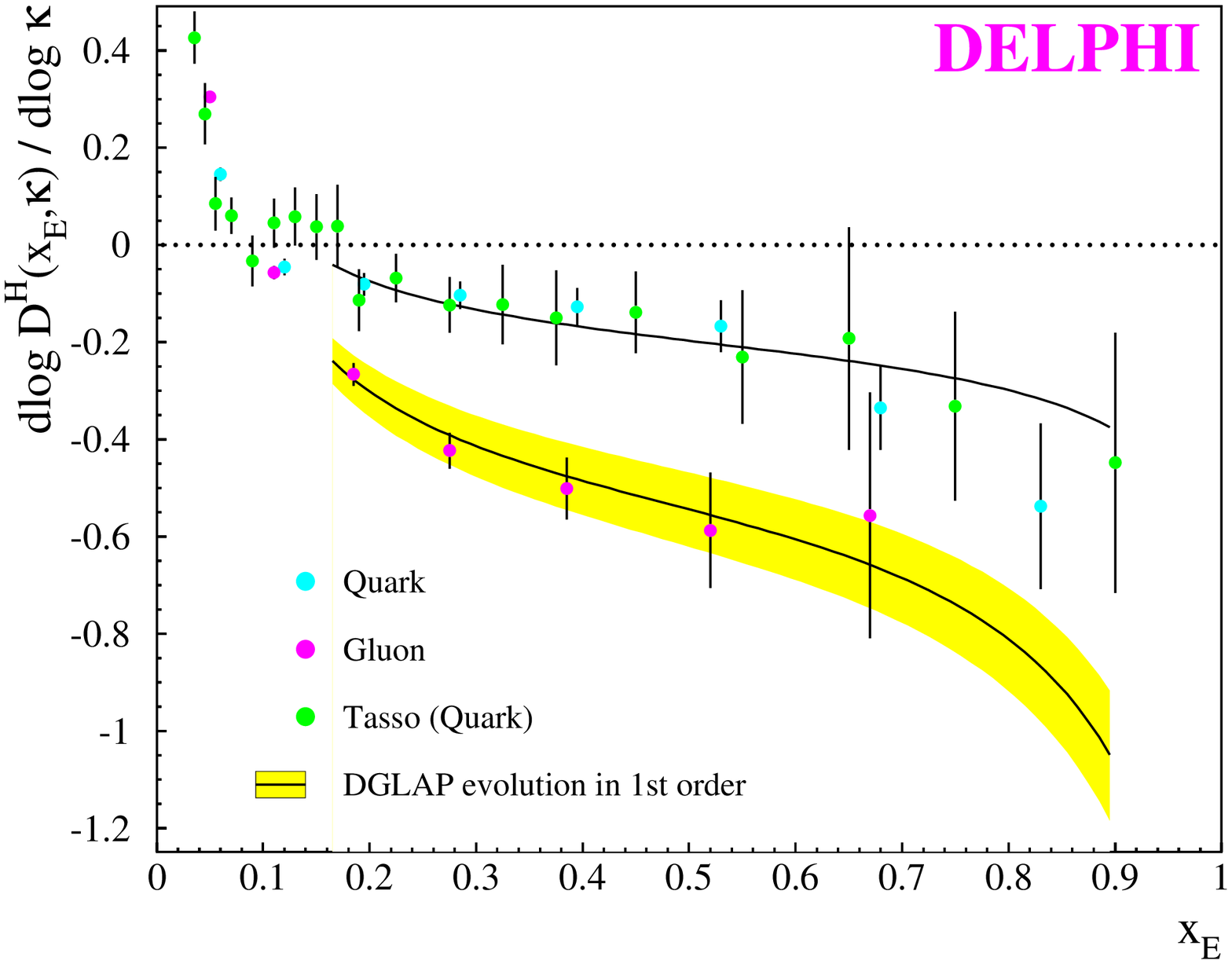}}
\caption{(Logarithmic) slopes of the scale ($\kappa_H$)
dependence of the quark and gluon
\ff s as a function of $x_E$}
\label{f:fanrat-upper}
\end{minipage}
\end{figure}

\fref{f:fan-low} shows
the measurement for light quark jets in \tje s as a function of $\kappa_H$,
represented by the open
symbols and superimposed  by  power law fits.
The corresponding result for gluons is shown in~\fref{f:fan-g-2}
for symmetric and asymmetric event
topologies.
The good agreement between the two event samples in~\fref{f:fan-g-2}
demonstrates that
$\kappa_H$ is a sensible choice for the scale of gluon jets as well.

A simultaneous fit with the first order DGLAP equations
to both the quark and gluon \ff s has been performed. Beyond the parameters
of the analytic ansatz of the \ff s , $\Lambda_{QCD}$ and $C_A$ have been
treated as free parameters. The fit is sensitive to the
occurrence of $C_A$ in the $g\to gg$ splitting kernel.
The fit describes the data well and yields:

\wi 0.48\textwidth
\fwi 0.46\textwidth

\[
\frac{C_A}{C_F}=2.44 \pm 0.21_{stat} {\rm (preliminary)} \, .
\]
for the colour factor ratio.

In \fref{f:fanrat-upper} the logarithmic slope of the gluon and quark
fragmentation functions are compared in dependence of $x_E$ superimposed by
the result of the DGLAP fits. The shaded
area indicates the statistical uncertainty of $C_A$.
The data points were obtained from power law fits to each $x_E$
interval individually.
As expected from the structure of the DGLAP equation, the scaling
violations are $\sim$ 2 times larger for gluons than for quarks.
Furthermore, the observed scaling violation in quark jets is in very good
agreement with the measurements of the {\sc Tasso}
Collaboration\lref{l:frag_tasso} (already visible in~\fref{f:fan-low}).

\subsection{Scale dependence of the gluon and quark jet \mul ies}
\label{sec:jetmul}

\begin{figure}[t]
\begin{minipage}[t]{\wi}
\centerline{\epsfxsize \fwi \epsfbox{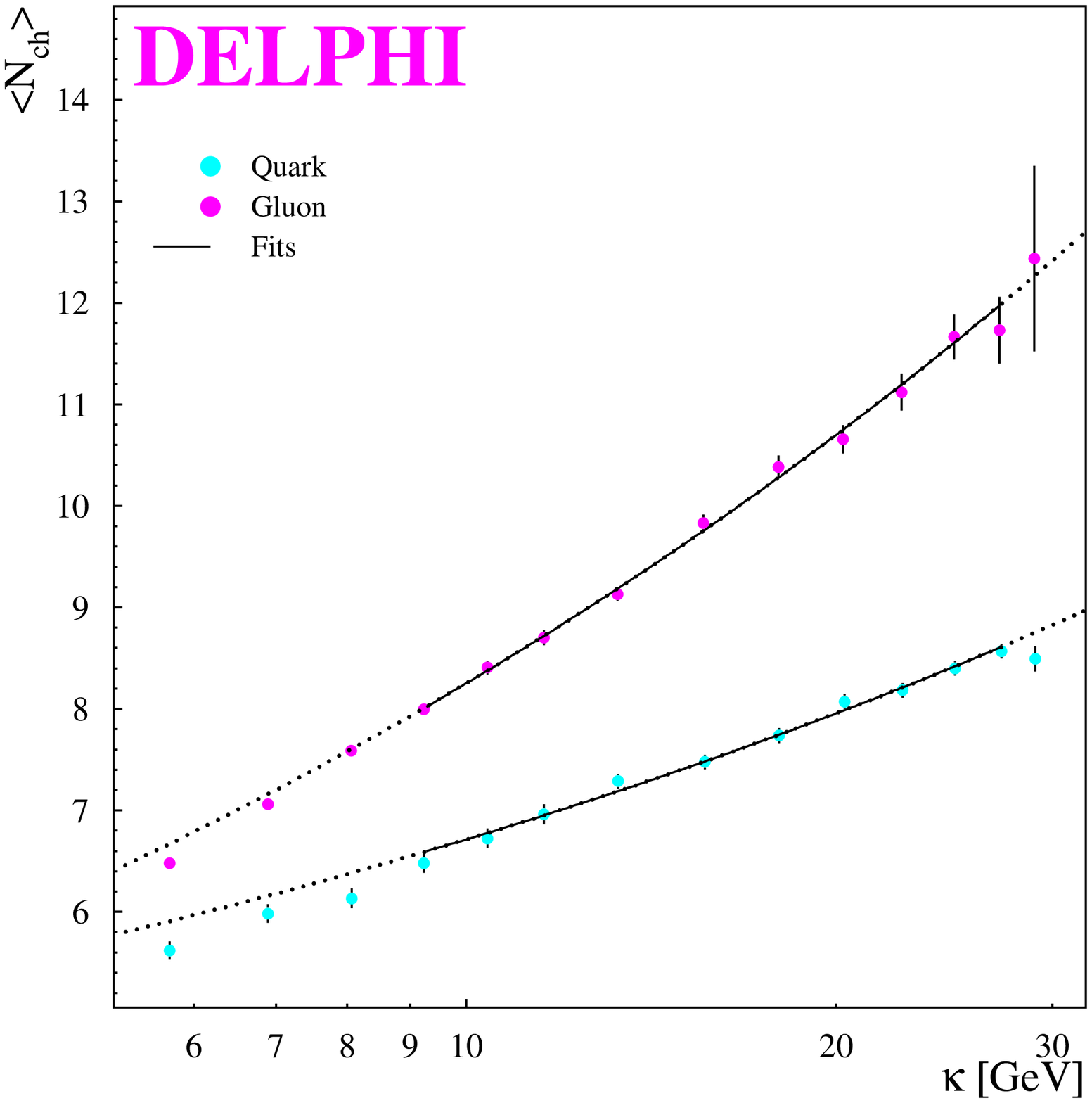}}
\caption{Average charged multiplicity for light quark and gluon jets
         as a function of $\kappa_H$ fitted with Eqn.\ref{eq:nq} and 
         \ref{eq:ng} }
\label{f:multi-fit}
\end{minipage}
\hfill
\begin{minipage}[t]{\wi}
 \centerline{\epsfxsize 9cm \epsfbox{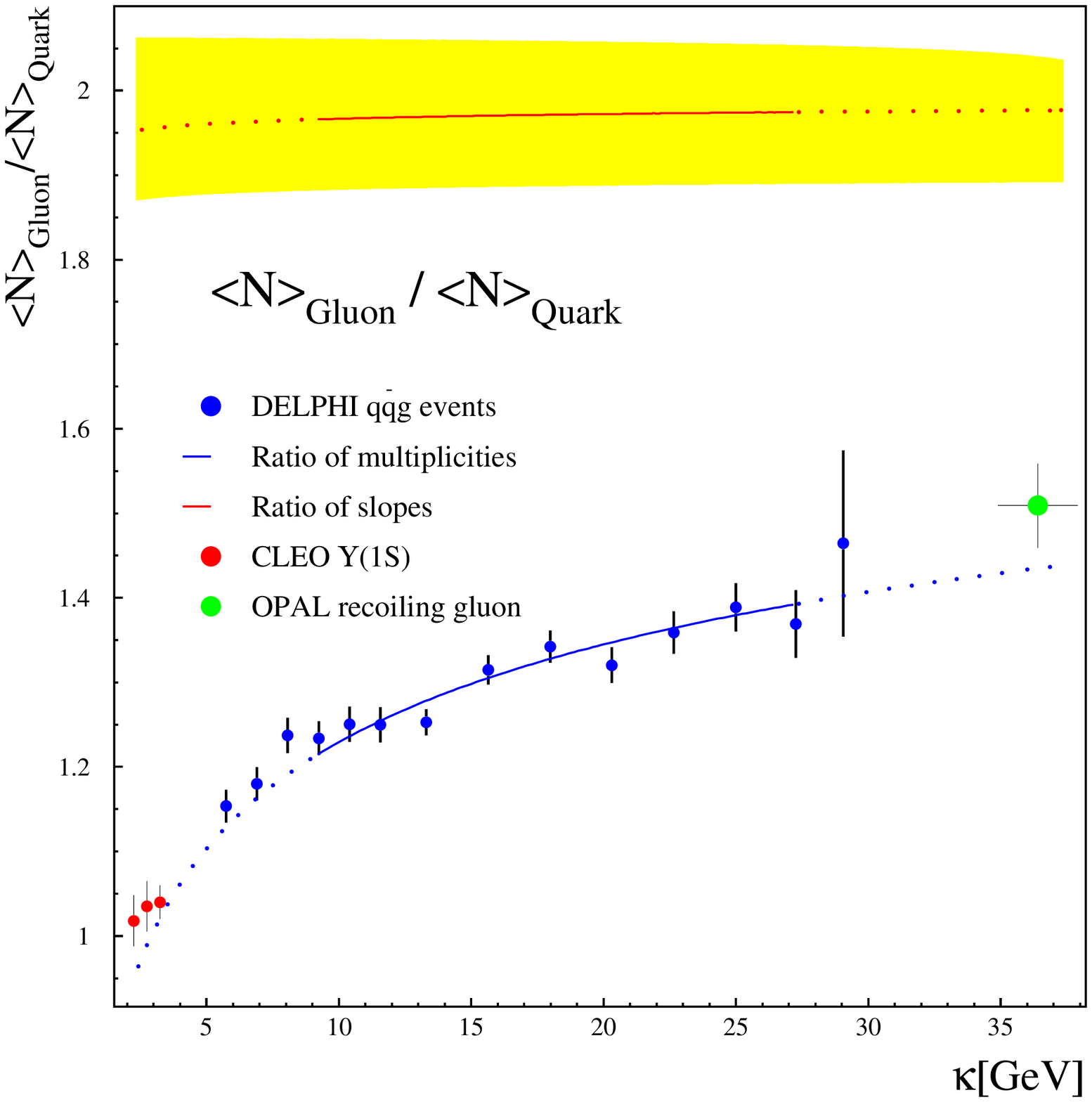}}
 \caption{Ratio of gluon and quark jet \mul ies and their slopes as a 
          function of $\kappa_H$}
 \label{f:multi-rat}
\end{minipage}
\end{figure}

There is a long standing QCD prediction that in the limit of
large parton (or jet)
energies
the ratio $r_{n}=\mean{N_g}/\mean{N_q}$ of the gluon over the quark
multiplicity should be equal to
\cacf\lref{l:brodsky_gunion}.
Following a NNLO calculation\lref{l:gaffney_mueller}, $r_{n}$ is reduced
by $\sim 10\%$ and 
nearly independent of energy.
\fref{f:multi-fit} shows the scale dependence of the jet multiplicity
and \fref{f:multi-rat} shows this ratio as a function of the hardness
$\kappa_H$.
Obviously, the ratio is much lower than expected; consequently
neither the LO nor the NNLO prediction for $r_{n}$ can describe the measured
data.

To describe the quark and gluon jet multiplicities, one would
set $\mean{N_q} = \mean{N_{pert}}$ as calculated
from some QCD approach\lref{l:DKT,l:webber} and
$\mean{N_g} = r_{n}\cdot\mean{N_{pert}}$, with 
$r(\kappa_H)=C_A/C_F[1-r_1\gamma_0(\alpha_S)-r_2\gamma_0^2(\alpha_S)]$.
A modification of this ansatz had to be performed
to obtain a sensible description of the measured data:
\begin{eqnarray}
\langle N_q\rangle (\kappa_H) & = & \langle N_{pert}\rangle (\kappa_H) + N_0^q
\label{eq:nq}\\
\langle N_g\rangle (\kappa_H) & = & \langle N_{pert}\rangle (\kappa_H)\cdot r_n + N_0^g
\label{eq:ng}
\end{eqnarray}
The introduction of the phenomenological offsets $N_0^q$ and $N_0^g$ 
which are assumed to be constant, accounts for expected differences of the
fragmentation of the leading quark or gluon.

\wi 0.48\textwidth
\fwi 0.46\textwidth

The quality of the fit increases from
$\chi^2/ndf$=14.5 to 0.85 when introducing these additional terms.
The difference
$N_0^q$-$N_0^g$ yields about 2,
consistent with about one additional {\em instable} primary particle
built in the primary quark fragmentation.
Unlike the multiplicity ratio itself, the ratio of derivatives of the gluon and
quark \mul ies with scale
$
\partial\mean{N_g}(\kappa_H)/\partial \kappa_H~/~
\partial \mean{N_q}(\kappa_H)/\partial \kappa_H
$
is in the predicted range of $\sim 2$ already at very small scale values
(see~\fref{f:multi-rat}). This
corresponds to the expectation that if the \mul y ratio is \cacf\ in the limit
of very large energies, the ratio of the slopes shows the same
behaviour, simply due to de l'H\^opital's rule. There should be
less sensitivity of the slope ratio to non-perturbative effects than
of the \mul y ratio itself.
Furthermore, the fit to the quark and gluon jet \mul ies extrapolates very well
to the \mul y ratios measured by {\sc Cleo} (direct measurement of
$\epem\to\Upsilon(1s)\to gg\gamma$ decays)\lref{l:mul:cleo} and
{\sc Opal} (analysis of most energetic jets)\lref{l:mul:opal},
which is a confirmation of this analysis.

Fitting~\eref{eq:nq}~and~\ref{eq:ng} to the data
yields $\cacf=2.12\pm0.10_{stat}$,
in agreement with QCD.
Nevertheless, large systematic errors are here to be expected for this value
due to ambiguities in the assignment of particles to
jets, the definition of the three jet region and the choice of the underlying
jet scale.

\subsection{Measurement of \cacf\ from \tje\ \mul ies}
\label{sec:khoze}

The uncertainties in the determination of the colour factor ratio quoted
in \chref{sec:jetmul} can be avoided by applying a MLLA
prediction\lref{l:khoze_ochs}
for \tje\ \mul ies, $N_{\qqbar g}$, to the data:
\begin{eqnarray}
N_{\qqbar g} &=& N_{\epem}(2E^*) + r_n(p_1^T)\cdot\left\{\frac{1}{2}\cdot
N_{\epem}(p_1^T) - N_0\right\}  \label{eqn:3mul_0} \\
{\rm with:~~} p_1^T & = & \sqrt{2
    \frac{ (p_qp_g) (p_{\bar{q}}p_g) }
         {p_qp_{\bar{q}} } }
~~~;~~~
2E^*  =  \sqrt{2p_qp_{\bar{q}}} \nonumber
~~~;~~~ p_i : {\rm four-momenta}
\end{eqnarray}
The first term represents the \mul y of an
\epem\ event at a CMS energy of the invariant mass of the $\qqbar$ system,
the second term
half the \mul y of a hypothetical $\epem\to gg$ event with a CMS energy of
twice the transverse momentum of the gluon.
Coherence effects are included by the exact definition of the scales.
Again, the term ``-$N_0$'' was introduced into the original MLLA prediction to
absorb non-perturbative contributions.

\setcounter{figure}{9}
\begin{floatingfigure}[l]{10cm}
 \centerline{\epsfxsize 10cm \epsfbox{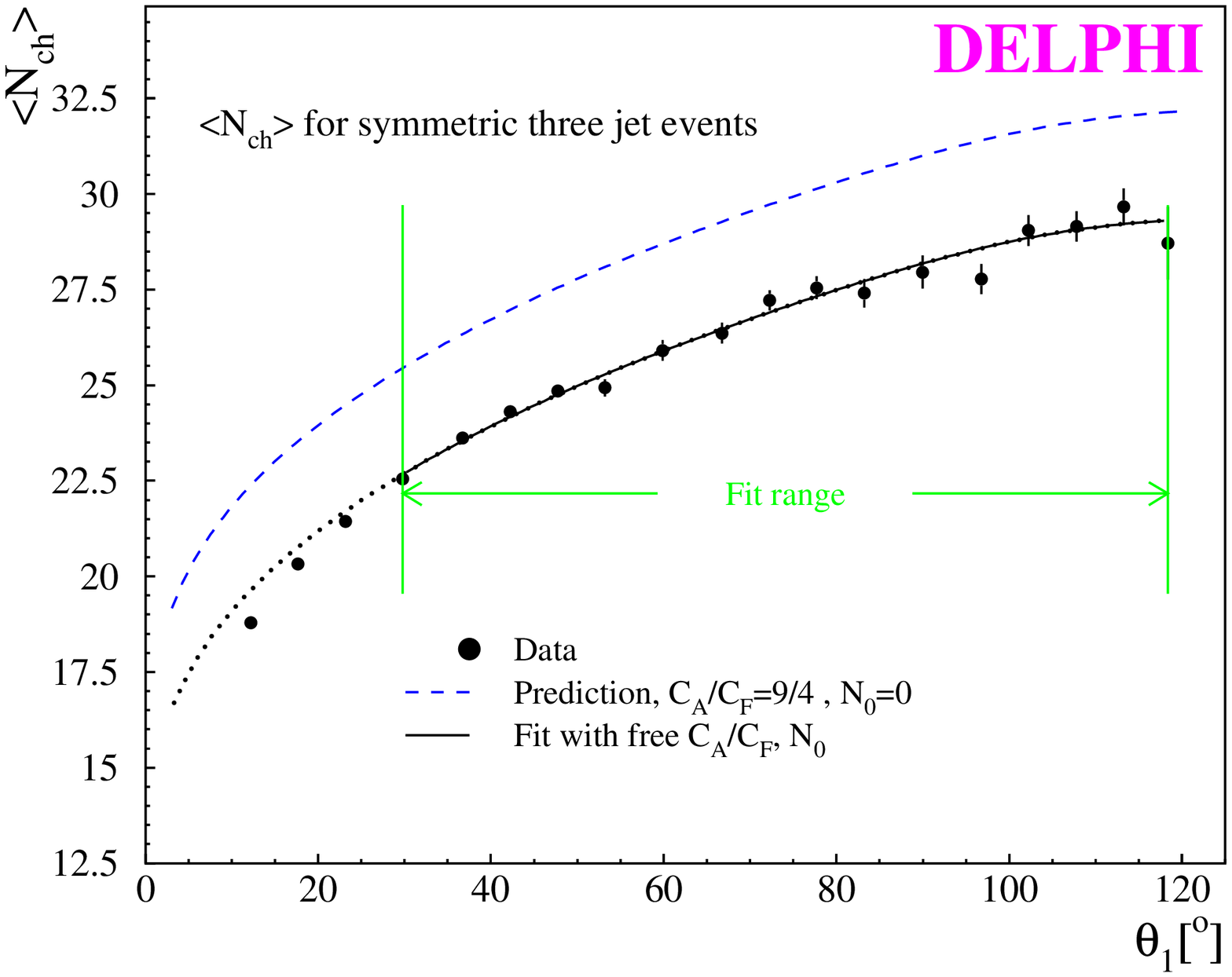}}
\centerline{
\parbox[b]{10cm}{
 FIG. 9. Three jet event \mul ies
 }
}
\end{floatingfigure}

\eref{eqn:3mul_0} has been applied to strictly symmetric events with
\mbox{$\theta_2,\theta_3 = (\pi-\theta_1/2) \pm 1.5^\circ$}.
Therefore the only parameter describing the
event topology is the opening angle $\theta_1$.
$N_{\epem}(\sqrt{s})$ has been chosen as 
$N_{pert}(\sqrt{s})$\lref{l:DKT,l:webber}
(as in~\chref{sec:jetmul});
the free parameters of these formulae are obtained from
the \mul ies of
\epem events as a function of the CMS energy.
{\sc Fig.~}8
shows the \tje\ \mul ies as a function of the opening angle
$\theta_1$. Superimposed is the fit with \cacf\ and $N_0$ as free parameters
as a solid line, extrapolated to the region contaminated by two jet events
($\theta_1<35^\circ$), which is excluded from the fit.
The dashed line represents the original prediction (\cacf =2.25, and $N_0=0$).
The need for some kind of correction to account for
non-perturbative effects is obvious, and the ansatz of adding a constant term
is a simple but effective choice.
The fit yields a very precise accurate measurement of \cacf\ :
\begin{eqnarray}
\frac{C_A}{C_F} &=& 2.246 \pm 0.062_{stat}\pm 0.080_{sys}\pm 0.095_{theo} \, .
\end{eqnarray}

\begin{figure}[t]
\wi 0.48\textwidth
\fwi 0.46\textwidth
\begin{minipage}[t]{\wi}
 \centerline{\epsfxsize \fwi \epsfbox{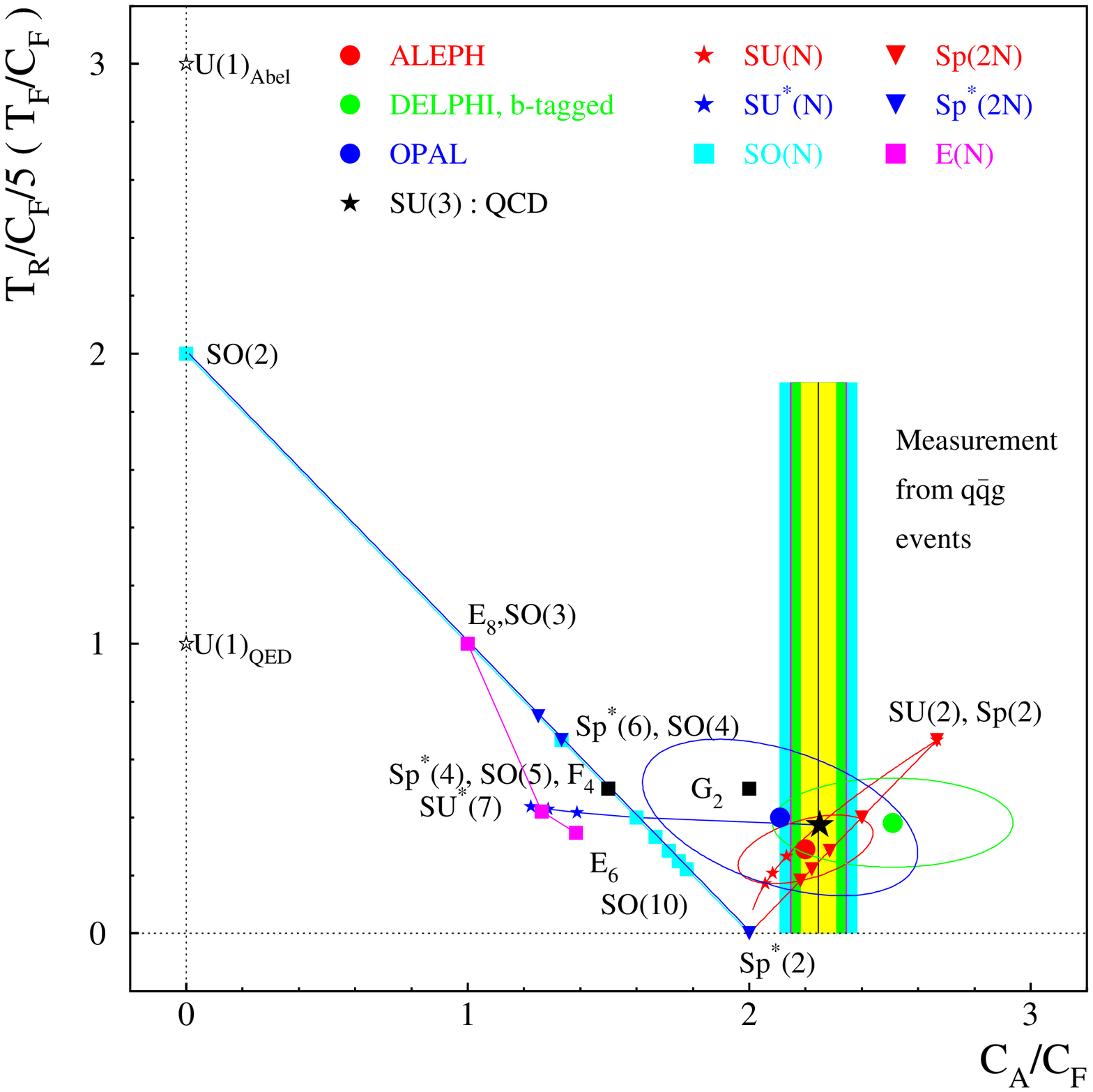}}
 \caption{The colour factor group plot}
 \label{f:kh-wim11-full}
\end{minipage}
\hfill
\begin{minipage}[t]{\wi}
\centerline{\epsfxsize \fwi \epsfbox{ 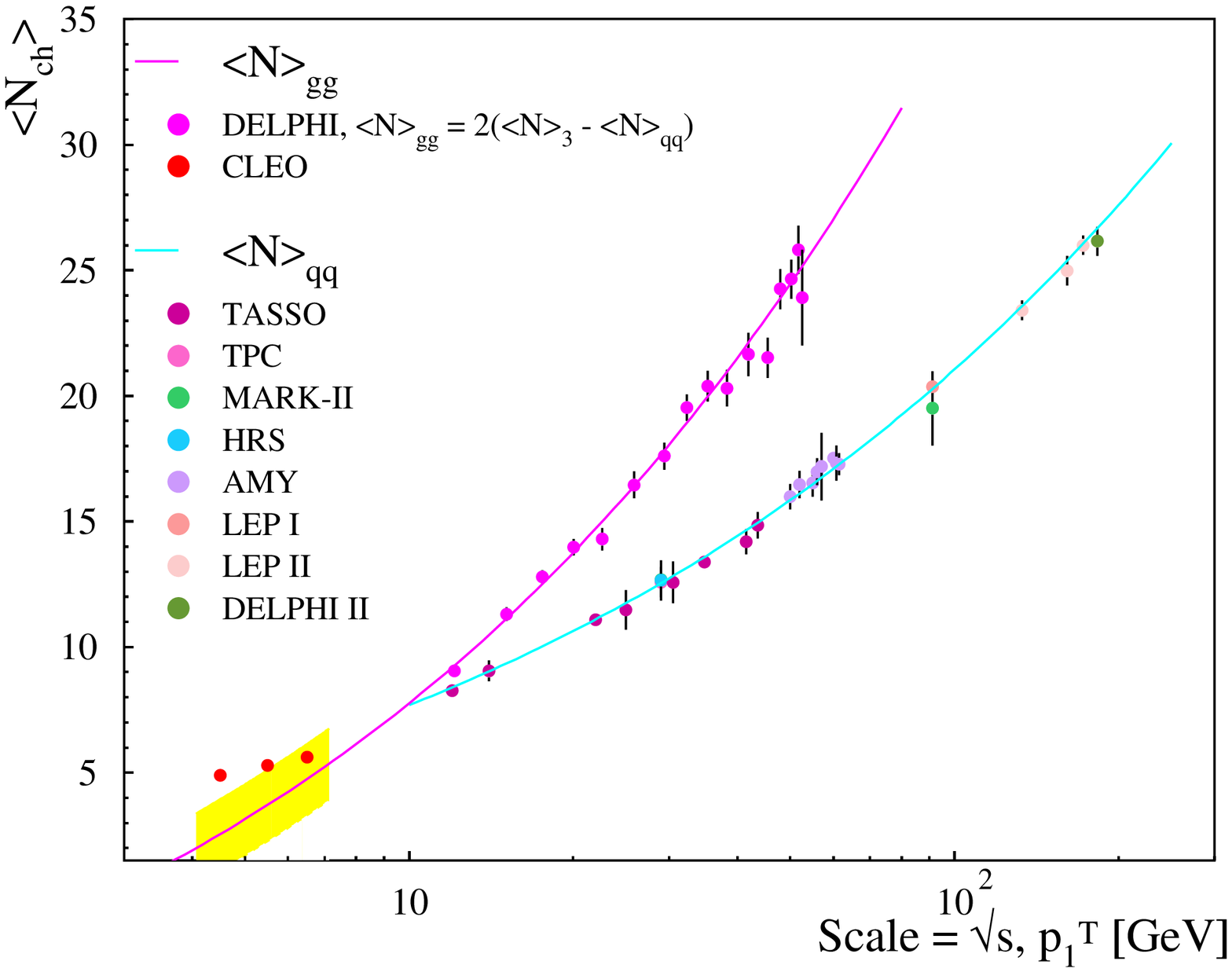}}
\caption{Comparison of the charged hadron multiplicity for an initial
\qqbar\ and a $gg$ pair as
function of the scale.
The dashed curve is a fit according of $N_{pert}(\sqrt{S})$,
the full line is twice the second term of
Eqn.~\ref{eqn:3mul_0}. The shaded band indicates the uncertainty due to
the error of $N_0$.}
\label{qq_gg_mult}
\end{minipage}
\end{figure}

The prediction of the multiplicity ratio given in
\cite{dremin_nechaitilo} has been tried as an alternative to the given $r_n$.
Although this calculation takes recoil effects into account,
a non-perturbative offset term is still required.
The prediction differs by about 10\% from \cite{l:gaffney_mueller} 
in the NNLO term.
As it does not reproduce the colour factor ratio contained in the
fragmentation models which describe the data well,
it has not been applied in this analysis.
\fref{f:kh-wim11-full} shows the obtained result for the colour factor ratio
in comparison of those obtained from four jet analyses. As far as we know
our result is the most accurate measurement of $C_A/C_F$ so far.

In order to illustrate comprehensively the contents of the measurement
of the three jet multiplicity we compare in Fig. \ref{qq_gg_mult} the
multiplicity corresponding to a $gg$ and a \qqbar\ final state.
The \qqbar\ multiplicity is taken to be the multiplicity measured in
\epem\ annihilation corrected for the ${\rm b\bar{b}}$ contribution.
The $gg$ multiplicity at low scale values is taken from the CLEO measurement
\cite{l:mul:cleo}, for which no systematic error was specified.
At higher scale, twice the difference of the three jet multiplicity and the
\qqbar\ term (the first term in Eqn.~\ref{eqn:3mul_0}) is interpreted as
the $gg$ multiplicity.
The dashed curve through the \qqbar\ points is a fit of $N_{pert}(\sqrt{S})$.
The $gg$ line is the perturbative expectation for back-to-back gluons
according to the second term of Eqn.~\ref{eqn:3mul_0}.
The plot shows again that the increase of the
$gg$ multiplicity with scale is about twice as big as in the \qqbar\ case,
illustrating the large gluon-to-quark colour factor ratio \cacf.

\section*{Acknowledgments}

I would like to express my gratitude to J. Drees for the possibility to join
the DPF '99. I thank
K.~Hamacher and M.~Siebel for the pleasant 
corporation in the gluon
analysis task of the Wuppertal group of \delphi .
Further I would like to acknowledge the work of the organizing committee
which enabled a very pleasant conference. In particular I thank
L.~Dixon and J.~Huston for the smooth organization of the session in which I
presented my talk.

%
%


\end{document}